\documentclass[a4paper]{jpconf}
\usepackage{graphicx}

\usepackage{hyperref}
\usepackage{amssymb}
\usepackage{longtable}
\usepackage{rotating}
\usepackage{color}
\usepackage{epsfig}
\usepackage{epsf}
\usepackage{fancyhdr}

\def\be{\begin{equation}}
\def\ee{\end{equation}}

\def\bea{\begin{eqnarray}}
\def\eea{\end{eqnarray}}

\begin{document}

\title[Recent results on the search for continuous sources with LIGO]
     {Recent results on the search for continuous sources with LIGO and GEO~600}

\author{Alicia M Sintes for the LIGO Scientific Collaboration}

\address{ Departament de F\'{\i}sica, Universitat de les Illes
Balears, Cra. Valldemossa Km. 7.5, E-07122 Palma de Mallorca,
Spain}

\address{ Max-Planck-Institut f\"ur
    Gravitationsphysik, Albert Einstein Institut, Am M\"uhlenberg 1,
    D-14476 Golm, Germany}

\ead{sintes@aei.mpg.de}

\begin{abstract}
An overview of the searches for continuous 
gravitational wave signals in LIGO and GEO~600 
performed on  different recent science runs and results are  presented. 
This includes both  searching for gravitational  waves from known pulsars
as well as blind searches over a wide parameter space.
\end{abstract}

\section{Introduction}

Construction of the LIGO \cite{ligo1, ligo2} and GEO~600 \cite{GEO1}
instruments began in the mid-1990s.
When the construction phase of the project was completed, the LIGO
instruments were officially inaugurated on November  1999.
Since then the commissioning of the instruments has alternated with a sequence of short
engineering and science runs  at increasing
sensitivity. The four science runs to date are:
S1: August 23 -- September 9, 2002,
S2: February 14 -- April 14, 2003,
S3: October 31, 2003 -- January 9, 2004 and
S4: February 22 -- March 23, 2005.
%
Both LIGO and GEO~600 are to begin a full science run (S5) in November, 
with the aim of
gathering data continuously for 18 months.
During the previous science runs although these instruments
were still to reach their design sensitivity, their performances were
sufficiently good to justify a serious test of our search algorithms on real
interferometer data, in particular, to search for continuous gravitational
waves from very dense, rapidly-spinning stars, such as neutron or quark stars.

Rapidly rotating neutron stars are the most likely sources of periodic,
persistent gravitational waves in the frequency band between
100 and 1000~Hz.
These objects may generate gravitational waves through a variety of mechanisms,
including non-axisymmetric distortions of the star, velocity
perturbations in the star's fluid, and free precession.
Regardless of the specific mechanism,
the emitted signal is a quasi-periodic continuous wave whose frequency
changes slowly during the observation time due to the intrinsic
frequency drift induced by the energy loss through gravitational wave emission
(and possibly other mechanisms) and the motion of the detector with respect
to the source. As the
intrinsic amplitude of gravitational waves from this class of sources
is several orders of magnitudes smaller than the typical
root-mean-square value of the noise, detection can only be achieved by means
of long integration times, of the order of weeks-to-months.

Code to search for this kind of sources has been under development within
the LIGO Scientific collaboration (LSC) since the mid- to late 1990s.
For S1--S4  the LSC continuous-wave search group has developed several
methods to search and set upper limits  on signals from radio
pulsars as well as to perform an all-sky search for unknown neutron stars
\cite{S1-CW, cw-prl, houghS2cqg, houghS2, fdsS2, timeS3, eh}.
So far none of the searches conducted  provided a detection but upper limits
were set on the gravitational wave emission and these
 results are summarized in this paper.

\section{Search methods and results}

In the S1 analysis \cite{S1-CW}, two techniques were used to set upper limits on
gravitational wave emission from pulsar J1939+2134 (the fastest rotating known
millisecond pulsar): a Bayesian time-domain method \cite{time,re} 
and a classical
frequency-domain method. The main result from this S1 analysis was an upper
limit on signals from pulsar   J1939+2134  of $h_0 <1.4\times 10^{-22}$ with
$95\%$ confidence.

For the LIGO S2 run, the time-domain search (which is more suitable for targeted
sources)  was expanded to include all well-known isolated pulsars  with putative
gravitational wave frequencies above 40 Hz \cite{cw-prl}. Using the S2 data,
multi-detector upper limits were set on gravitational wave emission from 28
pulsars including  J1939+2134  and the Crab pulsar. The tightest limit on
gravitational wave strain came from pulsar J1910$-$5959D with a $95\%$
upper limit that $h_0 <1.7\times 10^{-24}$. At that time this was the lowest
upper limit, for an astrophysical source, ever set by an interferometric
gravitational wave detector. The same method is currently
applied to analyze S3--S4 LIGO and GEO data \cite{timeS3}. The main change in
this search since S2 has been the addition of pulsars in binary systems.
Currently 93 pulsars are being searched of which 60 are binaries and 33 are
isolated. The improved sensitivity of the detectors in S3--S4 promise to give
interesting results for several sources. For the Crab pulsar, we should be
within a factor of a few  of the spin-down based upper limit. Moreover, for S5,
expectations for the Crab would be that within a year we would beat the
spin-down limit.

The frequency-domain statistical technique used in the S1 analysis  is
currently being used for broad all-sky searches for unknown sources
and for a rotating neutron star in a binary system.
In fact, this search has been the test-bench for the
core science analysis that the \textit{Einstein@home} \cite{eh} project
is carrying out. \textit{Einstein@home} is a
public distributed computing project that the LSC has been operating since
February 2005, built using the Berkeley Open Infrastructure for
Network Computing (BOINC).  Members of the general public can 
easily install the software on their 
personal computer.  When otherwise idle, their computer downloads data
from {\it Einstein@home}, searches it for pulsar signals, then uploads
information about any candidates.
Another example of the coherent frequency-domain technique is 
\cite{fdsS2} which uses S2 data in coincidence from two of
the LIGO detectors to perform  two different searches:
(i) for signals from isolated sources over the whole sky and the frequency band
160--728.8~Hz using 10 h of data, and (ii) for a signal from the Low-Mass X-ray
Binary Scorpius X-1 over orbital parameters in the frequency bands
 464--484~Hz and 604--624~Hz using 6 h of data. 

Future continuous wave searches will involve searching longer data stretches
(on the order of months to years) for unknown sources over a large 
 parameter space. It is well known that
the computational cost of coherent techniques for searches of this type is
absolutely prohibitive, thus hierarchical methods have been proposed
\cite{BCCS,bc,cgk,pss01}, where coherent and incoherent search stages
are alternated  in
order to identify  efficiently statistically significant candidates.
An essential step towards
the actual implementation of a ``hierarchical pipeline'' for production
analysis is the thorough investigation and characterization of its building
blocks -- the coherent and incoherent stages -- over a large parameter
space and on actual data sets; in fact the optimal sensitivity can
be ultimately achieved through careful tuning of
a variety of search parameters that are difficult to determine on pure
theoretical grounds, including the choice of thresholds at each stage,
the different tilings of the parameter space, the quality cuts in the data
and the choice of coincidence windows.

In~\cite{houghS2} we report results obtained by applying for the first
time  an incoherent analysis
to the data collected during the S2 run. The search method is based on the Hough
transform, which is a computationally efficient and robust pattern recognition
technique. We apply this technique  to perform an all-sky search for isolated
spinning neutron stars using the two months of data. The main results of this
paper are all-sky upper limits on the strength of gravitational waves emitted by
unknown isolated neutron stars on a set of narrow frequency bands in the range
200--400 Hz. Our best    $95\%$  frequentist upper limit  that we obtain in this
frequency range is  $h_0 <4.43\times 10^{-23}$. 
 Based on the statistics of
neutron star population with optimistic assumptions, this upper limit is about 1
order of magnitude larger than the amplitude of the strongest expected signal,
but with  1 yr of data at design sensitivity for initial LIGO, we should gain
about 1 order of magnitude in sensitivity, thus enabling us  to detect signals
smaller that what is predicted by the statistical argument mentioned above.

Other incoherent techniques, such as  ``stack-slide'' \cite{bc}
or  ``power-flux''
as well as the Hough transform \cite{hough04}, are used by
the LSC continuous-wave search group
 to analyze S4 data. All of them
use, in some way, the power from the Fourier transforms of short stretches of
data, which are added in a way that compensates for the Earth's motion and the
pulsar's spin-down during the observation period.
 These analyzes will provide us with the
first thorough understanding and characterization of such approaches and
allow us to place upper-limits on
regions of the parameter space that have never been explored before.

Future searches, using sophisticated hierarchical analysis techniques, 
 together with the computing power
  of {\it Einstein@home} will allow the deepest pulsar searches, and 
  thus initial LIGO at full sensitivity will have some chance of observing 
  a continuous gravitational wave signal.

\ack

The authors gratefully acknowledge the support of the United States National
 Science
 Foundation for the construction and operation of the LIGO Laboratory and the
 Particle Physics and Astronomy Research Council of the United Kingdom, 
 the Max-Planck-Society and the State of Niedersachsen/Germany for support 
 of the construction and operation of the GEO600 detector. The authors 
 also gratefully acknowledge the support
  of the research by these agencies and by the Australian Research Council, 
  the Natural Sciences and Engineering Research Council of Canada, 
  the Council of Scientific
 and Industrial Research of India, the Department of Science and Technology 
 of India, the Spanish Ministerio de Educaci\'on y Ciencia, 
 the John Simon Guggenheim Foundation,
 the David and Lucile Packard Foundation, the Research Corporation,
 and the Alfred P. Sloan Foundation.

\end{document}